\documentclass[prb,twocolumn,superscriptaddress,showpacs,amsmath,amssymb]{revtex4-1}

\usepackage{graphicx}
\usepackage{latexsym}
\usepackage{amsmath}
\usepackage{amssymb}
\usepackage{amsfonts}
\usepackage{changes}
\usepackage{color}
\usepackage{bm}
\usepackage{verbatim}
\usepackage{pagecolor,lipsum}
\usepackage{lineno}
\usepackage[version=3]{mhchem}

\usepackage{framed}
\definecolor{shadecolor}{RGB}{222,222,221}
\begin{document}

\title{Magnetic eigenmodes in chains of coupled $\varphi_0$-Josephson  junctions with ferromagnetic weak links}

\author{G.A. Bobkov}
\affiliation{Moscow Institute of Physics and Technology, Dolgoprudny, 141700 Russia}

\author{I. V. Bobkova}
\affiliation{Moscow Institute of Physics and Technology, Dolgoprudny, 141700 Russia}
\affiliation{National Research University Higher School of Economics, Moscow, 101000 Russia}

\author{A. M. Bobkov}
\affiliation{Moscow Institute of Physics and Technology, Dolgoprudny, 141700 Russia}

\date{\today}


\begin{abstract}
A coupled chain of superconductor/ferromagnet/superconductor (S/F/S) with anomalous ground phase shift $\varphi_0$ represents a system realizing long-range interaction between magnetic moments of the weak links. The interaction is of magnetoelectric origin and is mediated by the condensate phase of superconductors. The system is a paradigmic platform  for investigation of collective magnetic states governed by the superconducting phase. Here we study the magnetic eigenmodes of such a system and demonstrate that the eigenfrequencies are determined by the magnetic configuration of the whole system and are controlled by the superconducting phase. Depending on the orientation of the magnetic easy axis the eigenmodes can be very different  ranging from individual oscillations of different magnets to highly-cooperative behavior.
\end{abstract}

  \maketitle

{\it Introduction.} By now it is well established that by hybridizing superconducting and ferromagnetic orders intriguing physics emerges
and new device functionality can be achieved, which
is inaccessible in conventional systems. The first direction, which is frequently called superconducting spintronics \cite{Linder2015,Eschrig2015}, involves exploiting of proximity effects in heterostructures to obtain nontrivial superconducting spin states and their further use in spintronics and superconducting electronics. There are a lot of well-known examples: $\pi$-junctions \cite{Buzdin1982,Ryazanov2001,Yamashita2020,Feofanov2010}, memory elements \cite{Golovchanskiy2016,Vernik2013,Karelina2021,Guarcello2020}, superconducting spin valves \cite{Tagirov1999,Li2013,Gingrich2016,Lenk2017}, spin-caloritronics devices \cite{Bobkova2021,Huertas-Hernando2002,Giazotto2013,Giazotto2006,Giazotto2008,Giazotto2006review,Kawabata2013,Giazotto2015_2,Heikkila2018,Geng2023}.
 
The other and more recent direction can be called superconducting magnonics. It studies influence of a superconducting subsystem on spin excitations, in particular, the ability to modify and tune a spinwave dispersion and to manipulate with eigenstates of collective spin excitations via their interaction with a superconducting
subsystem. In particular, it was found that the adjacent metal works as
a spin sink strongly influencing Gilbert damping of the
magnon modes \cite{Tserkovnyak2005,Ohnuma2014,Bell2008,Jeon2019,Jeon2019_2,Jeon2019_3,Jeon2018,Yao2018,Li2018,Golovchanskiy2020,Silaev2020}. 
It was also found that adjacent superconducting
layers can result in shifting of $k = 0$ magnon frequencies
\cite{Jeon2019_3,Li2018,Golovchanskiy2020,Silaev2020,Silaev2022,Golovchanskiy2023}. Nontrivial modifications of spectra of different magnetic excitations caused by an interaction with superconducting subsystem were reported. The interaction can be both of electromagnetic origin and due to proximity effects. The electromagnetic interaction between the ferromagnet and superconductor results in the appearance of skyrmion-fluxon excitations \cite{Hals2016,Baumard2019,Dahir2019,Menezes2019,Andriyakhina2021,Petrovic2021,Bihlmayer2021}, magnon-fluxon excitations \cite{Dobrovolskiy2019} and efficient gating of magnons \cite{Yu2022}. The interaction via the proximity effect in thin film ferromagnet/superconductor(F/S)
and antiferromagnet/superconductor(AF/S) hybrids results in the appearance of composite particles, composed of a magnon in F(AF) and an accompanying cloud of spinful triplet pairs in S, which were termed magnon-cooparon \cite{Bobkova2022,Bobkov2023}. 

Another interesting type of interaction between the magnetic and superconducting subsystem can be realized under the presence of inversion symmetry breaking via the magnetoelectric effect, which provides a direct coupling between a supercurrent or a phase of superconducting condensate and a magnetic moment, thus opening up broad prospects for various options for controlling magnetization using supercurrent \cite{Bobkova_review}. 

The most well-studied realization of the magnetoelectric coupling is the so-called $\varphi_0$-Josephson junction \cite{Krive2004,Nesterov2016,Reynoso2008,Buzdin2008,Zazunov2009,Brunetti2013,Yokoyama2014,Bergeret2015,Campagnano2015,Konschelle2015,Kuzmanovski2016,Malshukov2010,Tanaka2009,Linder2010,Zyuzin2016,Lu2015,Dolcini2015}. Under the simultaneous breaking of inversion symmetry and a time-reversal symmetry, a supercurrent can be induced in the Josephson junction (JJ) at zero phase difference between the leads.  In the ground state of
the junction this ”anomalous supercurrent” is compensated by the phase shift $\varphi_0 \neq 0,\pi$, which is called the anomalous ground state phase shift. The breaking of the inversion symmetry leads to the appearance of spin-orbit coupling.  The breaking of the time-reversal symmetry can be achieved more easily by applying a magnetic field to the JJ. For this reason Josephson junctions with anomalous phase shift have already been realized experimentally by several groups in semiconducting nanowires, quantum wells with strong Rashba spin-orbit coupling and on the surface of a topological insulator under the Zeeman effect of the applied magnetic field \cite{Mayer2020,Szombati2016,Assouline2019,Murani2017}. The time reversal symmetry can be also broken in a more involved way, for example, by creating an imbalance in the tunneling rates for spin up and spin down electrons of the edge states of 2D topological insulator (2D TI) \cite{Vigliotti2023}.

The other way to break the time reversal symmetry is to exploit ferromagnetic interlayers. It is more challenging, but interesting because opens rich possibilities to study the effects of the magnetoelectric coupling between the magnetic moment of the weak link and the condensate phase.   To realize this coupling one can use for the interlayers 2D or quasi 2D ferromagnets, where the Rashba spin-orbit coupling can be strong due to the structural inversion symmetry breaking. The other way is to exploit ferromagnetic insulator/3D topological insulator hybrids as interlayers \cite{Chang2013,Kou2013,Kou2013_2,Chang2015,Jiang2014,Wei2013,Jiang2015,Jiang2016}.  The possibility to obtain in such systems a direct coupling between the magnetization of the weak link and the superconducting phase opens great perspectives for applications of such structures for controlling magnetization \cite{Konschelle2009,Shukrinov2017,Nashaat2019,Rabinovich2019,Guarcello2020,Bobkova2020,Bobkova_review}.

\begin{figure}[tb]
	\begin{center}		\includegraphics[width=75mm]{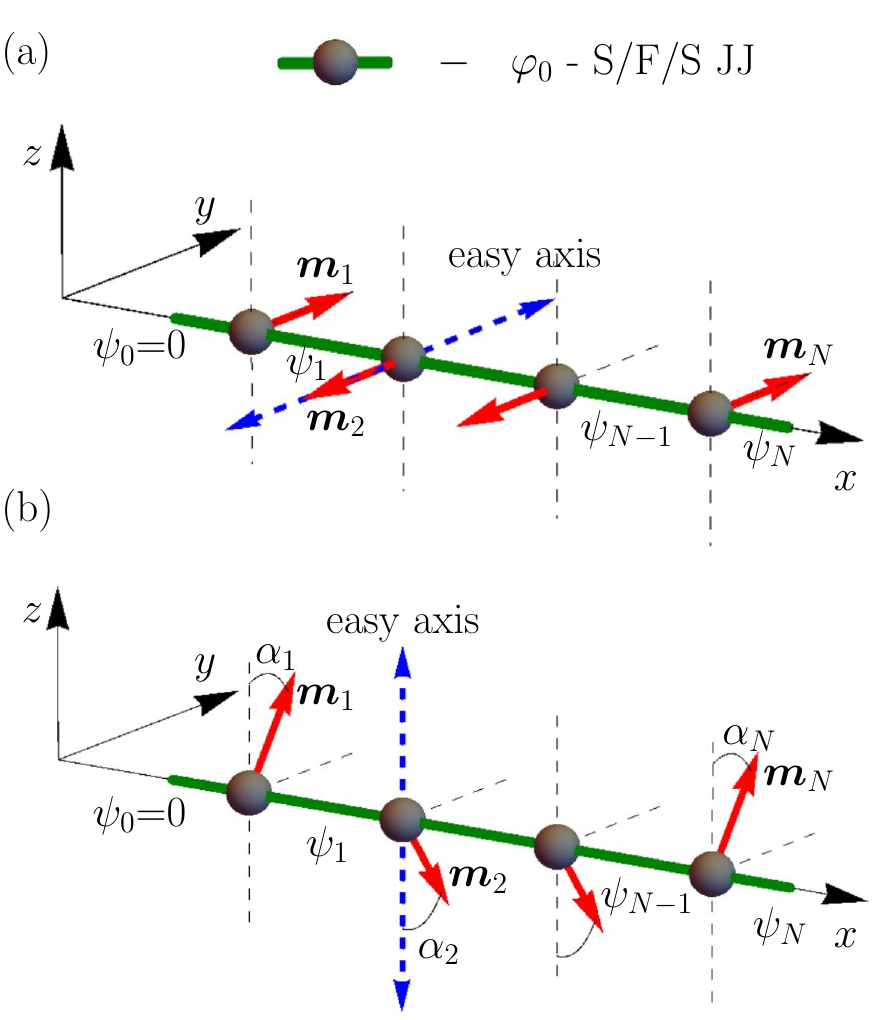}
		\caption{Sketch of the system under consideration. Ferromagnetic weak links (grey balls) are connected via superconductors (green lines). Magnetic moment of each weak link is shown by a red arrow. $\psi_i$ is a phase of the $i$-th superconductor in the chain. (a) Magnetic easy axis (dashed blue line) is along the $y$-axis, that is aligned with the direction of the spin-orbit induced effective field, see text. In equilibrium all the magnetic moments are directed along the easy axis. (b) Magnetic easy axis is along the $z$-axis, that is perpendicular to the direction of the spin-orbit induced effective field. In equilibrium the magnetic moments make some angles with the easy axis. The angles depend on the external phase difference $\psi_N$. }
  \label{fig:sketch}
	\end{center}
 \end{figure}

Further it was demonstrated \cite{Bobkov2022,BobkovG2023} that if the $\varphi_0$ - S/F/S JJs are coupled into a Josephson chain, see Fig.~\ref{fig:sketch}, there is a long-range interaction between them. Estimates made in Ref.~\onlinecite{Bobkov2022} suggest that the interaction strength is not reduced considerably even at the macroscopic scales of the order of millimeters. At larger distances $l$ between the magnets the coupling constant exhibits the long-range power law $1/l$ behavior. Further it was shown that such a chain manifests properties of $n$-level system, where the energies of the levels are only determined by projections of the total magnetic moment $\sum \bm M_i$, where $\bm M_i$ is a magnetic moment of $i$-th JJ, onto the easy magnetic axis. It resembles an atom in a Zeeman field, but the role of the field is played by the magnetoelectric coupling. It is interesting that, unlike the Zeeman effects in the atom, the relative order of energies of different states is controlled by phase difference between the external superconducting leads.

In this work we study eigen magnetic excitations in a system of magnets, which are weak link of coupled $\varphi_0$ - S/F/S  JJs and interact via the superconducting condensate. It is shown that due to the long-range interaction between the magnets the energies of the eigenmodes can depend on the equilibrium magnetic configuration of all the magnets. The eigenfrequencies are controlled via the external superconducting phase. The eigenvectors of these modes, that is the distribution of oscillations of different magnets for a given mode, are also investigated.

{\it System and model.} We consider a linear chain of $N$ coupled $\varphi_0$ - S/F/S JJs, where S means a conventional superconductor and F means a homogeneous ferromagnet with magnetic moment $\bm M_i$, where $i$ is a number of the weak link in the chain. Further we introduce unit vectors along the direction of the corresponding magnetization $\bm m_i = \bm M_i/|\bm M_i|$. It is assumed that the ferromagnets are easy-axis magnets. The mutual orientation of the easy axis and the spin-orbit induced effective field (see below, directed along the $y$-axis) determines behavior of the eigenmodes. For this reason in this work we consider two different orientations of the easy-axis: along  the $y$-direction and along the $z$-direction, see Fig.~\ref{fig:sketch}. As it is demonstrated below, they realize two opposite limiting cases of the magnetoelectric behavior. The superconducting phase difference $\psi_N$ between the leads is an external controlling parameter. 

The current-phase relation (CPR) of a separate S/F/S junction takes the form $I = I_{c,i} \sin (\chi_i-\varphi_{0,i})$, where $I_{c,i}$ is the critical current of $i$-th magnet, $\chi_i = \psi_i - \psi_{i-1}$ is the superconducting phase difference at this JJ and $\varphi_{0,i}$ is the anomalous phase shift for a given JJ. In general, anomalous phase shift $\varphi_{0,i}$ in S/F/S JJs depends on the direction of magnetization $\bm m_i$. The particular form of this dependence is determined by the type of spin-orbit coupling (SOC). For definiteness we consider Rashba type SOC because it arises due to the structural inversion asymmetry and is the most common type of SOC for low-dimentional ferromagnets and thin-film ferromagnet/normal metal (F/N) hybrid structures. In this case the anomalous phase shift takes the form 
\begin{eqnarray}
\varphi_0 =  r \hat {\bm  j} \cdot (\bm n \times \bm m ),
\label{phi_0}
\end{eqnarray}
where $\hat {\bm  j}$ is the unit vector along the Josephson current and $\bm n$ is the unit vector describing the direction of the structural asymmetry in the system. For the case under consideration it is along the $z$-axis. $r$ is a constant quantifying the strength of the coupling between the magnetic moment and the condensate. It depends on the material parameters of the ferromagnet, Rashba constant $\alpha$, length of the ferromagnetic interlayer and was calculated in the framework of different models, including ballistic Rashba ferromagnets, diffusive Rashba ferromagnets and S/F/S JJs on top of the 3D TI \cite{Buzdin2008,Bergeret2015,Zyuzin2016,Nashaat2019}. See Ref.~\onlinecite{Bobkov2022} for more details on the particular expressions of the magnetoelectric coupling constant $r$ in the framework of particular models.   The results presented below depend only on the symmetry of Eq.~(\ref{phi_0}) expressing how the anomalous phase shift depends on the direction of the magnetization $\bm m_i$, and the dependence of the constant $r$ on the junction parameters is irrelevant for our conclusions.
If we choose $x$-axis along the Josephson current, then symmetry of our system dictates that
\begin{eqnarray}
\varphi_{0,i} = r_i m_{yi}.
\label{chi0}
\end{eqnarray}
This relation also survives in the dynamic situation $\bm m_i = \bm m_i(t)$ and has been used  for calculation of the magnetization dynamics in voltage-biased and current-biased JJs \cite{Nashaat2019,Konschelle2009,Shukrinov2017,Guarcello2020}. 

In the framework of our model we assume that the critical current $I_{c,i}$ does not depend on the direction of the magnetization $\bm m_i$. In fact, the behavior of the critical current depends crucially on the particular type of the considered S/F/S JJ. For example, it can be independent on the magnetization direction, as it has been reported for the ferromagnets with SOC \cite{Buzdin2008}, or it can depend strongly on  the $x$-component of the magnetization, as it takes place for the ferromagnetic interlayers on top of the 3D TI \cite{Zyuzin2016,Nashaat2019}. The influence of this dependence on the magnetic modes of the system will be considered elsewhere.

The energy of the system consists of Josephson energies of  all junctions and easy-axis anisotropy energies of  all magnets:
\begin{eqnarray}
E =  \sum \limits_{i=1}^N \Bigl[ \frac{\hbar I_{c,i}}{2e}\bigl(1-\cos(\psi_i-\psi_{i-1}-\varphi_{0,i})\bigr) - \nonumber \\ 
\frac{K_i V_{F,i}}{2} (\bm m_{i} \bm e_i)^2 \Bigr],
\label{energy}
\end{eqnarray}
where the first term is the Josephson energy $E_J$ and the second term is the magnetic anisotropy energy $E_M$. $\bm e_i$ is the unit vector along the easy axis, $K_i$ - is the anisotropy constant of $i$-th magnet and $V_{F,i}$ is its volume.
$\psi_{i}$ is a phase of $i$-th superconductor (see  Fig.~\ref{fig:sketch}). 

In the static situation, when oscillations of magnetic moments are not excited, the superconducting phases at each of the superconductors can be found from the condition of the conservation of the supercurrent $I_{chain}$ flowing via the Josephson chain:  
\begin{align}
I_{chain} = &I_{c,i}\sin(\psi_i-\psi_{i-1}-\varphi_{0,i}) = \nonumber \\
&I_{c,j}\sin(\psi_j-\psi_{j-1}-\varphi_{0,j}) 
\label{current_conservation}
\end{align}
for arbitrary $i$ and $j$. 
In general, one should take into account the phase gradient due to the supercurrent flowing through the system. This leads to the fact that the phase $\psi_i$ of $i$-th superconductor is not constant, $\psi_i(x) = \psi_{i,l} + \kappa_i (I_{chain}/I_{c,i})(x/L)$, where $\psi_{i,l}$ is the superconducting phase at the left end of $i$-th superconductor, $L$ is its length and the second term accounts for the phase gradient due to the supercurrent $I_{chain}$ flowing through the system. To simplify the analysis we disregard the order parameter phase gradient. Numerical values of $\kappa$ were estimated in Ref.~\onlinecite{Bobkov2022}  for realistic systems and it was concluded that the phase gradient can be safely neglected at least up to submillimeter lengths of the superconductors. In the present work we assume that all the coupled JJs are identical, that is they all have the same parameters $r$, $I_c$ and $K V_{F}$. 

To have controllable magnetoelectric coupling between the magnetic moments it is important to have a fixed phase difference $\psi_N$ between the external superconducting leads (not a fixed external current) \cite{Bobkov2022}. Experimentally the phase $\psi_N$ can be fixed and controlled by several ways. We assume that our system is inserted into an asymmetric Josephson interferometer, where it is in parallel with an ordinary Josephson junction with a much higher critical current. Then the total external current $I$ and the  phase $\psi_N$ between the external superconducting leads are related as
\begin{eqnarray}
I = I_{c,large}\sin \psi_N + I_{chain}, 
\label{current_psi}
\end{eqnarray}
where $I_{c,large}$ is the critical current of the additional ordinary Josephson junction. We assume that $I_{c,large} \gg I_{c,i}$. In this case $\psi_N \approx \arcsin[I/I_{c,large}]$.

 \begin{figure}[tb]
	\begin{center}		\includegraphics[width=75mm]{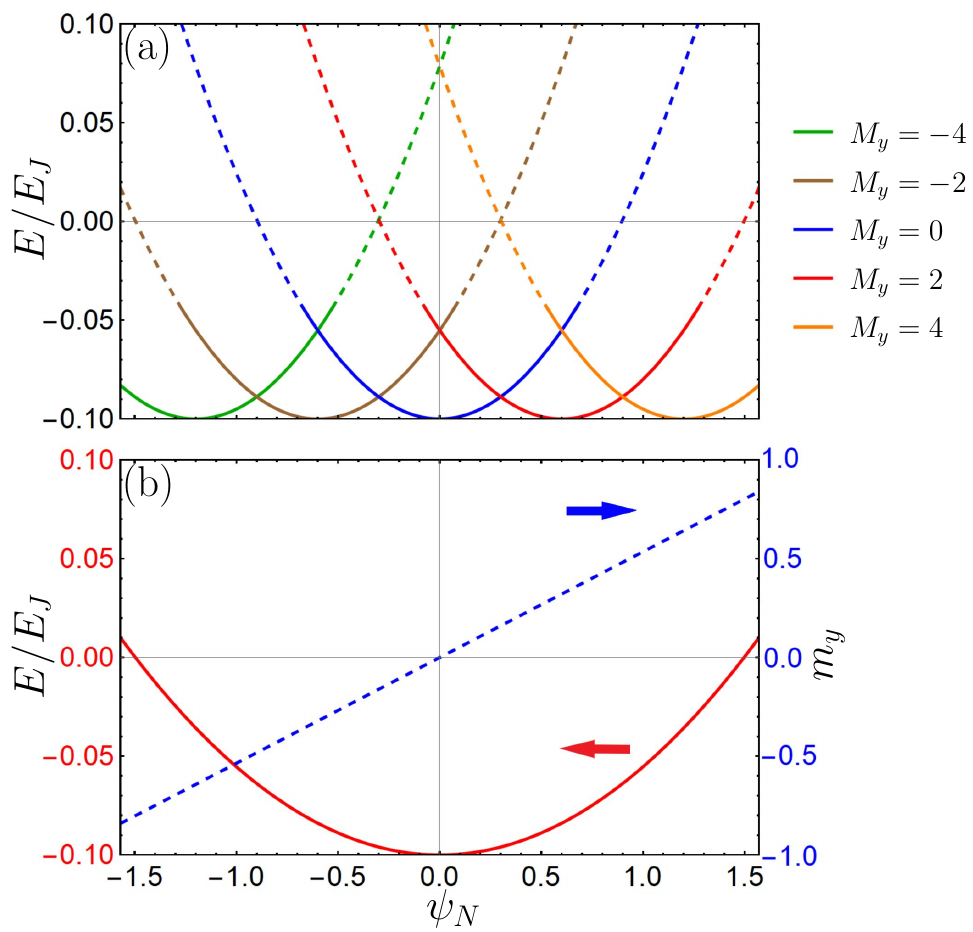}
		\caption{(a) Energy of stable magnetic configurations as a function of the external phase difference $\psi_N$. Easy axis is along the $y$-axis. Only the lowest branch for each of possible magnetic configurations, corresponding to different $y$-projections $M_y$ of the total magnetic moment, is plotted. Dashed parts of the curves represent unstable parts of the branches, where the corresponding magnetic configuration cannot exist and the system will spontaneously go to another possible magnetic configuration. (b) Easy axis is along the $z$-axis. Red curve: energy of stable magnetic configurations as a function of $\psi_N$. There is no energy splitting for different magnetic configurations. Blue curve: $m_y$ (the same for all magnetic moments) as a function of $\psi_N$. In different magnetic configurations individual magnetic moments can have different $m_{zi} = \pm \sqrt{1-m_y^2}$, but $m_y$ is the same for all the magnets and is determined by $\psi_N$, see Fig.~\ref{fig:sketch} for illustration. Parameters $N=4,~r=0.3,~E_M^0/E_J^0=0.025$ for the both panels.}
  \label{fig:energies}
	\end{center}
 \end{figure}

{\it Static magnetic configurations.} First of all we discuss static magnetic configurations, which are stable for a given external phase difference $\psi_N$. The case of easy-axis along the $y$-direction has already been considered in detail in Ref.~\onlinecite{BobkovG2023}. Here we just briefly remind the main results. The stable branches of the total energy of the system as a function of the external phase difference $\psi_N$ are decsribed by the following expression:
\begin{eqnarray}
E = N E_J^0 (1-\cos \Phi) - 
E_M^0 \sum \limits_{i=1}^N m_{yi}^2, 
\label{energy_1}
\end{eqnarray}
with $E_J^0 = \hbar I_c/2e$, $E_M^0 = K V_F/2$, and
\begin{eqnarray}
\Phi = \frac{\psi_N}{N} - \frac{\sum \limits_{i=1}^N \varphi_{0,i}}{N} + \frac{2\pi n}{N},
\label{Phi_sol}
\end{eqnarray}
where $n $ is an integer number. For a given magnetic configuration, that is for a given values $m_{yi}$ at all weak links, the total energy of the system $E(\psi_N)$, described by Eq.~(\ref{energy_1}), as a function of the phase difference between the external superconducting leads $\psi_N$ has $N$ different branches. Further we only consider the lowest branch for any magnetic configuration because it is quite difficult to place the system in higher states and keep it there in a controllable way. In general, stable magnetic configurations can be so-called "corner" magnetic configurations, when $m_{yi} =\pm 1$ for any $i$, and "non-aligned" magnetic configurations with $|m_{yi}| \neq 1$ can also be realized. It depends on the parameters $E_M^0/E_J^0$ and $r$. The full phase diagram of the static magnetic configurations has been studied in Ref.~\onlinecite{BobkovG2023}. Here we consider the parameter region, where only "corner states" are allowed. Then the lowest energy branches corresponding to different magnetic configurations are shown in Fig.~\ref{fig:energies}(a). It is seen that due to the nonzero magnetoelectric coupling $r \neq 0$ the degeneracy of the states corresponding to different magnetic configurations is removed. The states corresponding to different projections of the total magnetic moment on the $y$-axis $M_y = \sum \limits_{i=1}^N m_{yi}$  have different energies. It is also seen from Eq.~(\ref{Phi_sol}) because $\sum \limits_{i=1}^N \varphi_{0,i} \propto M_y$. 

The situation is completely different if the easy axis is along the $z$-direction. The total energy of the system is still determined by Eq.~(\ref{energy_1}) with the substitution $m_{yi} \to m_{zi}$:
\begin{eqnarray}
E = N E_J^0 (1-\cos \Phi) - 
E_M^0 \sum \limits_{i=1}^N m_{zi}^2, 
\label{energy_z}
\end{eqnarray}
In equilibrium a magnetic moment is aligned with the full effective field acting on the magnet. For the case under consideration the effective field consists of the magnetic anisotropy field $(K/M)m_{zi}\bm e_z$ and the   spin-orbit induced effective field (Rashba pseudomagnetic field) $\bm H_R \propto [\hat {\bm j} \times \bm n]$. Consequently, equilibrium magnetic configurations correspond to $m_{xi} = 0$, $m_{zi} = \pm \sqrt{1-m_{yi}^2}$. For the case of a separate $\varphi_0$ - S/F/S JJ such a reorientation of the easy axis by the Josephson current was reported in Ref.~\onlinecite{Shukrinov2018}. By numerical investigation of the total energy we found that at least for the chosen set of parameters the minimal energy is reached if all $m_{yi}$ have the same values $m_y$. Then from the condition $dE/dm_y = 0$ we find that in the stable magnetic configuration $m_y$ obeys the following equation:
\begin{eqnarray}
m_y = \frac{r E_J^0}{2E_M^0} \sin (\frac{\psi_N}{N} - r m_y).
\label{my_z_easy}
\end{eqnarray}
For the chosen set of parameters it has the only solution $m_y^0$. If $m_y^0 \in [-1,1]$, then it gives the $y$-component of all magnetic moments $m_{y,st}$. If $|m_y^0|>1$, then all the moments are directed along the  spin-orbit induced effective field $\bm H_R \propto [\hat {\bm j} \times \bm n]$, that is $|m_{y,st}| = 1$ and $\bm m \parallel \bm H_R$.  The stable magnetic configurations are described by $\bm m_i = (0, m_{y,st}, \pm \sqrt{1-m_{y,st}^2})^T$. All this magnetic configurations belong to the same energy, as it is seen from Eq.~(\ref{energy_z}). Therefore, in case if the easy axis is along the $z$-direction, that is perpendicular to the spin-orbit induced effective field, the magnetoelectric coupling does not remove the degeneracy of the total energy for different stable magnetic configurations. The dependence of the energy on $\psi_N$ is shown in Fig.~\ref{fig:energies}(b).

In this sense we can say that in the framework of the static problem of studying the stable magnetic configurations the strongest magnetoelectric splitting is realized if the magnetic easy axis is aligned with the Rashba pseudomagnetic field, and the splitting vanishes if the magnetic easy axis is perpendicular to the Rashba pseudomagnetic field. But it does not mean that for the perpendicular orientation the magnetoelectric impact on the system vanishes. Below we will see that if we consider the excitations of the system from a stable magnetic configuration, the physics is also very very sensitive to the orientation of the easy axis. The influence of the magnetoelectric coupling on the eigen modes is strong in the both cases, but the structure of the modes is very different.

{\it Magnetic eigenmodes.} The eigenfrequencies and eigenmodes of coupled system of magnetic moments $\bm m_i$ can be found in a quite standard way from the Landau-Lifshitz-Gilbert equation:
\begin{eqnarray}
\frac{\partial\bm m_i}{\partial t} = -\gamma \bm m_i \times \bm H_{eff,M}^i + \alpha \bm m_i \times \frac{\partial\bm m_i}{\partial t} - \nonumber \\
\frac{\gamma r I_{chain}}{2e M V_F}[\bm m \times \bm e_y],~~~~~~
\label{LLG}
\end{eqnarray}
where $\gamma$ is the gyromagnetic ratio,  $\bm H_{eff,M}^i = (K/M) (\bm m_i \bm e) \bm e$ is the local effective field in the ferromagnet induced by the easy-axis magnetic anisotropy and $\alpha$ is the Gilbert damping constant. The last term in Eq.~(\ref{LLG}) describes the spin-orbit torque, exerted on the magnet by the electric current $I_{chain}$  \cite{Yokoyama2011,Miron2010,Bobkova2018,Bobkova2020,Bobkov2022}. 
In the presence of magnetization dynamics the total current flowing through each of the JJs consists of the supercurrent and the normal quasiparticle current contributions \cite{Rabinovich2019}:
\begin{eqnarray}
I_{chain}=I_c \sin (\psi_i - \psi_{i-1} - \varphi_{0,i})+ \nonumber \\
\frac{1}{2eR_N}(\dot \psi_i - \dot \psi_{i-1} - \dot \varphi_{0,i}),
\label{current_total}
\end{eqnarray}
where $R_N$ is the normal state resistance of a separate S/F/S JJ. However, we consider a realistic case $e I_c R_N/\hbar \ll \gamma K/M$, which means that the dynamics of the superconducting phases is much faster than the dynamics of the magnetic subsystem. In this case we can assume that from the point of view of the magnetization dynamics all the superconducting phases are stationary, that is $\dot \psi_i = 0$, and neglect the second term in Eq.~(\ref{current_total}). Then the torque from the supercurrent can be accouted for via an additional contribution to $\bm H_{eff}$ as $\bm H_{eff,J} = -(1/M V_F)dE_J/d\bm m$ \cite{Nashaat2019,Konschelle2009,Shukrinov2017,Guarcello2020}. We have checked that our analytical results obtained in the framework of this assumption are in excellent agreement with the results of direct numerical calculations basing on Eqs.~(\ref{LLG}) - (\ref{current_total}).

If the easy axis is along the $y$-direction, the total effective field acting on each of the magnets $\bm H_{eff}^i = \bm H_{eff,M}^i+ \bm H_{eff,J}^i$ is also along the $y$-direction and takes the form
\begin{eqnarray}
\bm H_{eff}^i = \frac{K}{M}\left[ m_{yi} +   \frac{r E_J^0}{2 E_M^0}\sin \Phi\right] \bm e_y . 
\label{H_easy_y}
\end{eqnarray}
Substituting this effective field into the LLG equation
\begin{eqnarray}
\frac{\partial\bm m_i}{\partial t} = -\gamma \bm m_i \times \bm H_{eff}^i + \alpha \bm m_i \times \frac{\partial\bm m_i}{\partial t} 
\label{LLG_2}
\end{eqnarray}
and linearizing Eq.~(\ref{LLG_2}) with respect to the deviation $\delta \bm m_i$ of all the magnetic moments from the stable magnetic configuration $m_{yi} = \pm 1$, disregarding the Gilbert damping $\alpha \ll1$ we obtain the following eigenfrequencies:
\begin{eqnarray}
\omega_\pm =  \gamma \frac{K}{M} (1\pm \frac{r E_J^0}{2E_M^0}\sin \Phi) .
\label{eigen_frequency_y}
\end{eqnarray}
The Gilbert damping term gives the decay rate of the eigenfrequencies. In the first order with respect to $\alpha$ one obtains $\omega \approx \omega_\pm + i \alpha \omega_\pm$. It is seen that due to the presence of the magnetoelectric coupling $r$ the standard frequency of the ferromagnetic resonance (FMR) of an easy axis magnet $\omega_0 = \gamma K/M$ is split into two frequencies. The amplitude of the splitting depends on the particular stable magnetic configuration and on the external phase via $\Phi$. This is very different  from a conventional Heisenberg chain of $N$ magnets coupled by short-range exchange interactions, where we should obtain $N$ modes corresponding to different possible discrete wave vectors. 

 \begin{figure}[tb]
	\begin{center}		\includegraphics[width=75mm]{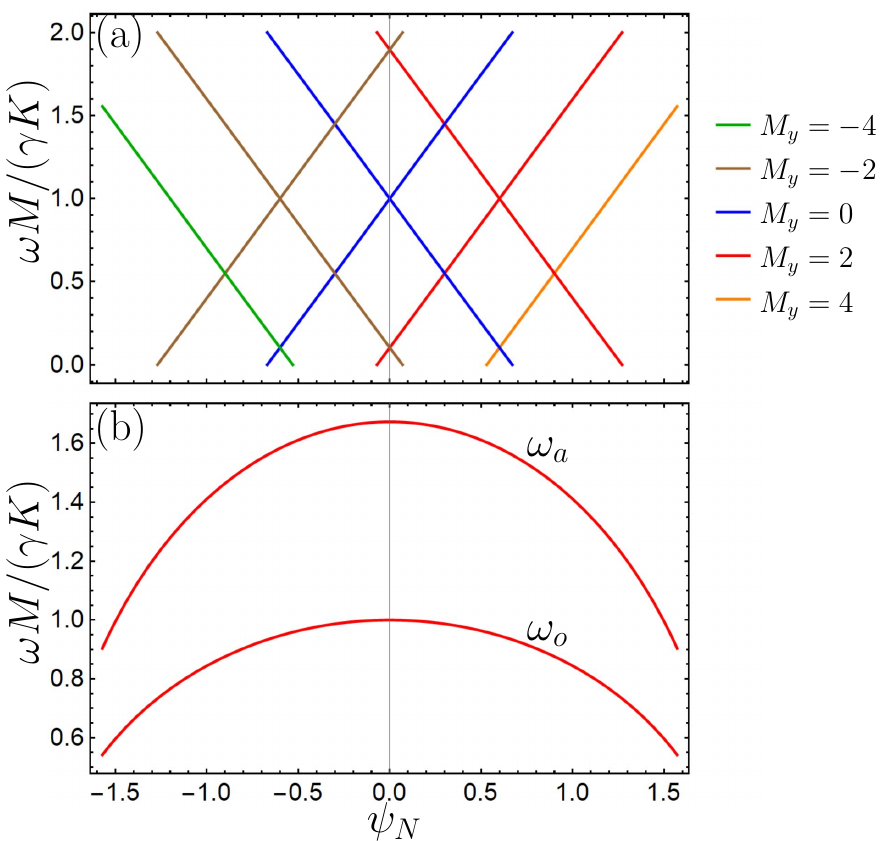}
		\caption{Eigenfrequencies of magnons as functions of the external phase $\psi_N$. (a) Magnetic easy axis is along the $y$-axis. There are two eigenfrequencies $\omega_\pm$ for a magnetic configuration corresponding to a given $M_y$. The modes only exist when the appropriate magnetic configuration is stable, see Fig.~\ref{fig:energies}(a). (b) Magnetic easy axis is along the $z$-axis. There are only two eigenfrequencies, acoustic mode $\omega_a$ and optic mode $\omega_o$, which are the same for all magnetic configurations. Parameters $N=4,~r=0.3,~E_M^0/E_J^0=0.025$ for the both panels.}
  \label{fig:frequencies}
	\end{center}
 \end{figure}

\begin{figure}[tb]
	\begin{center}		\includegraphics[width=75mm]{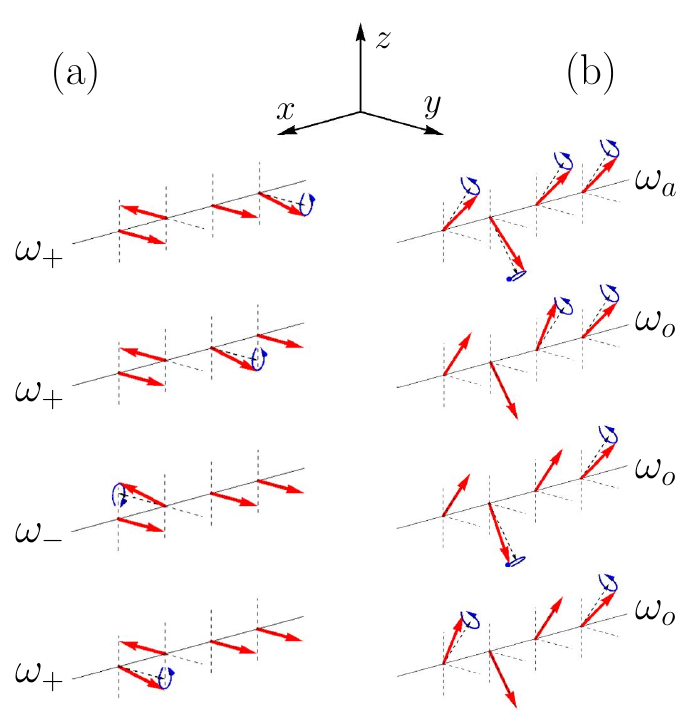}
		\caption{Schematic illustration of the structure of eigenmodes corresponding to eigenfrequencis plotted in fig.~\ref{fig:frequencies}. (a) Magnetic easy axis is along the $y$-axis. As an example the magnetic configuration $M_y=2$ is shown. Eigenfrequency $\omega_+$ is 3-fold degenerate. The corresponding eigenmodes can be chosen as independent oscillations of  magnets with $m_{yi} = 1$. Eigenfrequency $\omega_-$ is non-degenerate because only one of the magnets has $m_{yi} = -1$.  (b) Magnetic easy axis is along the $z$-axis. The acoustic eigenfrequency $\omega_a$ is non-degenerate. In this mode all the magnets rotate with the same $\delta m_y (t)$. The optic eigenfrequency $\omega_o$ is three-fold degenerate. The corresponding eigenmodes have zero total $y$-projection of the oscillation amplitude $\sum \limits_i \delta m_{yi}(t) = 0$.}
  \label{fig:modes}
	\end{center}
 \end{figure}

The eigenfrequencies $\omega_\pm$ are plotted in Fig.~\ref{fig:frequencies}(a) as functions of $\psi_N$. As it was found in Ref.~\onlinecite{BobkovG2023}, we can change stable magnetic configuration of the system by varying $\psi_N$. With such a change, the eigenfrequencies will experience a jump, as it is seen in Fig.~\ref{fig:frequencies}(a). The degree of degeneracy of the eigenfrequencies also depends on the magnetic configuration. Frequencies $\omega_\pm$ are $(N \pm M_y)/2$-fold degenerate. The eigenmodes, corresponding to a given eigenfrequency can be chosen as independent oscillations of separate magnets, see Fig.~\ref{fig:modes}(a).  $(N+M_y)/2$ magnets with $m_{yi} = 1$ can be excited separately with a resonant frequency $\omega_+$, while $(N-M_y)/2$ magnets with $m_{yi} = -1$ can be excited separately with a resonant frequency $\omega_-$. It is interesting that inspite of non-decaying character of the interaction between the weak links, one can excite an independent oscillations of a separate magnet at a resonant frequency, which is determined by the whole magnetic configuration.

  \begin{figure}[tb]
	\begin{center}		\includegraphics[width=70mm]{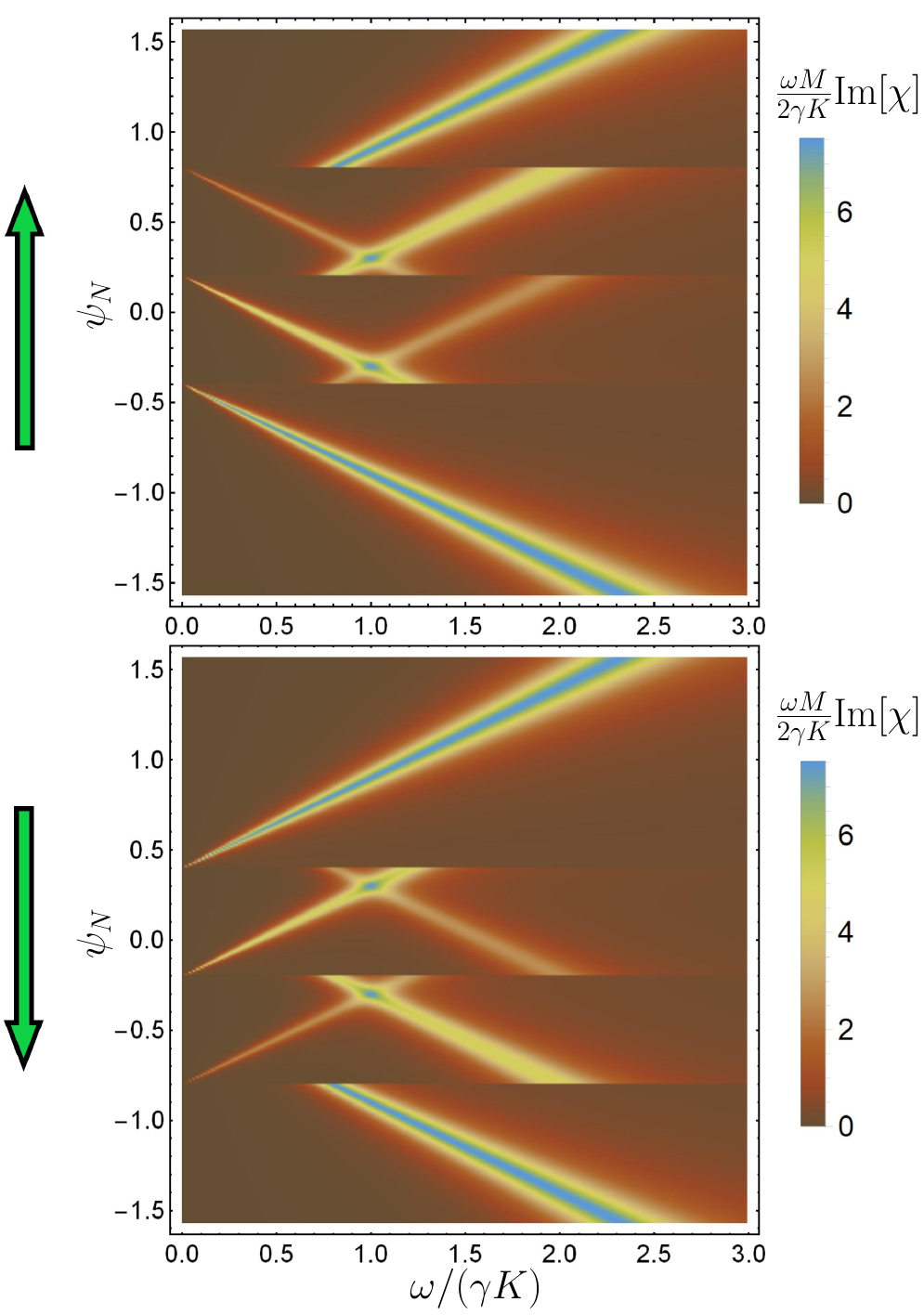}
		\caption{Absorption power as function of external frequency $P(\omega) \propto \frac{\omega}{2}{\rm Im}[\chi]$ upon adiabatic variation of $\psi_N$ in opposite directions $-\pi/2 \to \pi/2$ and $\pi/2 \to -\pi/2$ (shown by green arrows). Easy axis is along the $y$-direction. $N=3,~r=0.3,~E_M^0/E_J^0=0.025$.}
  \label{fig:hysteresis}
	\end{center}
 \end{figure}

The dependence of the eigenfrequencies on the external phase $\psi_N$ can be observed by varying the total current $I$ of the asymmetric Josephson interferometer adiabatically, see Eq.~(\ref{current_psi}). In this case it is possible to observe a hysteretic behavior of peaks in the absorption power, which correspond to the eigenfrequencies, upon varying the phase in opposite directions, as it is shown in Fig.~\ref{fig:hysteresis}. The hysteresis is due to the presence of several stable magnetic configurations for a given phase difference, which can be reached upon the adiabatic variations of the phase \cite{BobkovG2023}.

Now let us consider the case when the easy axis is along the $z$-axis. For each of the magnets we move to a local reference frame determined by three mutually orthogonal unit vectors $\bm e_x$, $\bm e_\parallel \equiv \bm m_i^0$, $\bm e_\perp = \bm e_\parallel \times \bm e_x$, where $\bm m_i^0$ is the unit vector along the equilibrium direction of a given magnet and $\delta \bm m_i$ is due to its excitation. We denote the angle between the vectors $\bm e_z$ and $\bm e_\parallel$ as $\alpha_i^0$. The angle between $\bm m_i = \bm m_i^0 + \delta \bm m_i$ and $\bm e_z$ is $\alpha_i = \alpha_i^0 + \delta \alpha_i$. Then we can write $\bm H_{eff}^i = H_{eff,\parallel}^i \bm e_\parallel + H_{eff,\perp}^i \bm e_\perp$ with
\begin{eqnarray}
H_{eff,\parallel}^i \approx \frac{K}{M} \left[\cos^2 \alpha_i^0 +\frac{r E_J^0}{2E_M} \sin \alpha_i^0 \sin \Phi \right]
\label{H_parallel]}
\end{eqnarray}
and
\begin{eqnarray}
H_{eff,\perp}^i \approx \lambda_0 \delta \alpha_i + \lambda_1 \cos \alpha_i^0 \sum \limits_{j=1}^N  \delta \alpha_j \cos \alpha_j^0  ,
\label{H_perp}
\end{eqnarray}
where $\Phi = (1/N)[\psi_N - r \sum \limits_{i=1}^N \sin \alpha_i^0]$, $\lambda_0 = (K/M) \sin^2 \alpha_i^0$ and $\lambda_1 = -(r^2 E_J^0/V_F M N) \cos \Phi \cos^2 \alpha_i^0$.

Since there are two stable solutions $\bm m_i^0 = (0, m_{y,st}, \pm \sqrt{1-m_{y,st}^2})^T$, we can write that $\sin \alpha_i^0 $ is the same for all magnets $\sin \alpha_i^0 \equiv \sin \alpha^0$ and  $\cos \alpha_i^0 = \sigma_i \cos \alpha^0$, where $\sigma_i = \pm 1$. Then the effective field $H_{eff,\parallel}^i \equiv H_{eff,\parallel}$ is also the same for all the magnets.  Expanding the excitation $\delta \bm m$ over two components $\delta \bm m = m_{i,\perp} \bm e_\perp + m_{i,x} \bm e_x$ and taking into account that $m_{i,\perp} = \delta \alpha_i$ we can write
\begin{eqnarray}
H_{eff,\perp}^i \approx \lambda_0 m_{i,\perp} + \lambda  \sum \limits_{j=1}^N  \sigma_i \sigma_j m_{j,\perp}  ,
\label{H_perp_2}
\end{eqnarray}
with $\lambda = \lambda_1 \cos^2 \alpha_i^0$. The linearized LLG equation takes the form:
\begin{align}
\dot m_{i,x} &= -\gamma (m_{i,\perp} H_{eff,\parallel}^i - H_{eff,\perp}^i) - \alpha \dot m_{i,\perp}, \nonumber \\
\dot m_{i,\perp} &= \gamma m_{i,x} H_{eff,\parallel}^i  + \alpha \dot m_{i,x}.
\label{LLG_lin}
\end{align}
Looking for the solution of Eqs.~(\ref{LLG_lin}) in the form $m_{i,x} = i X_i e^{i\omega t}$, $m_{i,\perp} = Y_i e^{i\omega t}$, we obtain the following equation for the eigenfrequencies and eigenvectors:
\begin{align}
\omega X_i &= \gamma H_{eff,\parallel} Y_i - \gamma (\lambda_0 Y_i + \lambda \sigma_i \sum \limits_{j=1}^N \sigma_j Y_j ) + i \alpha \omega Y_i,  \nonumber \\
\omega Y_i &= \gamma H_{eff,\parallel} X_i + i \alpha \omega X_i .
\label{eq_modes_z}
\end{align}
The resulting eigenfrequencies neglecting $\alpha $ take the form:
\begin{equation}
\omega_a = \gamma \sqrt{H_{eff, \parallel}(H_{eff, \parallel} - \lambda_0 - N \lambda)}, 
\label{frequency_1_z}
\end{equation}
\begin{equation}
\omega_o = \gamma \sqrt{H_{eff, \parallel}(H_{eff, \parallel} - \lambda_0 )}. 
\label{frequency_2_z}
\end{equation}
Taking into account the Gilbert damping results in $\gamma H_{eff,\parallel} \to \gamma H_{eff,\parallel} + i \alpha \omega$. Then the decay rate of the eigenmode $\omega_{a}$ is $\alpha (H_{eff,\parallel} - (\lambda_0+N\lambda)/2)$ and of the eigenmode $\omega_{o}$ is $\alpha (H_{eff,\parallel} - \lambda_0/2)$, respectively. The eigenfrequencies depend on the external phase difference $\psi_N$ via $\alpha^0$ and via the dependence on $\Phi$ in $\lambda_1$. The frequencies are plotted in Fig.~\ref{fig:frequencies}(b) as functions of $\psi_N$. Since in the case when the easy axis is along the $z$-axis there is no splitting of the equilibrium spectra for different $M_y$, in contrast to the case of easy $y$-axis, here we can see only one branch for each of the frequencies. The investigation of the eigenvectors indicates that the frequency $\omega_a$ is non-degenerate and the frequency $\omega_0$ is $(N-1)$-fold degenerate. 

The schematic illustration of the eigenmodes, corresponding to the eigenfrequencies, is presented in Fig.~\ref{fig:modes}(b). First of all, it is worth noting that, unlike the case of $y$-easy axis, the eigen modes cannot be chosen as independent oscillations of the magnets.  We see that $\omega_a$ corresponds to the motion of all the magnetic moments with the same values of the projection of the oscillation amplitude $\delta m_y(t)$ on the $y$-axis. Therefore, it can be interpreted as an acoustic mode. $\omega_o$ can be interpreted as an optic frequency corresponding to $N-1$ different modes with zero total angular $y$-projection of the oscillation amplitude $\sum \limits_i \delta m_{yi}(t) = 0$.

{\it Conclusions.} In conclusion, we have investigated magnetic eigenmodes in a system of magnets representing weak links of coupled $\varphi_0$ - S/F/S Josephson junctions. The interaction between the magnets is mediated by the superconducting condensate. It is long-range and can be assumed non-decaying in the considered range of distances between the magnets. The coupling between the magnetic moments and the condensate is of the magnetoelectric origin and produced via the interaction of the magnetic moment with the spin-orbit induced Rashba pseudomagnetic field, generated by the Josephson current, flowing through the chain of the JJs. Eigenfrequencies   are determined by the collective state of all the magnets and can be controlled by the superconducting phase between the external superconducting leads.  Possible stable states of the magnetic system crucially depend on the orientation of the magnetic easy  axis with respect the pseudomagnetic field. For this reason the eigenfrequencies and eigenmodes are very sensitive to the orientation of the magnetic easy  axis. 

If the easy axis and the pseudomagnetic field are aligned, the magnetoelectric splitting of different stable magnetic configurations is the most pronounced and determined by the projection of the total magnetic moment on the easy axis $M_y$. There are two eigenfrequencies for each $M_y$. The corresponding eigenmodes can be chosen diagonal in the basis of magnetizations of individual weak links. It is wonderful that the eigenfrequencies are determined by the projection of the total magnetization $M_y$, but the magnets behave like completely uncoupled. 

On the contrary, if the easy axis and the pseudomagnetic field are perpendicular, the magnetoelectric splitting of different magnetic configurations is absent. There are two eigenfrequencies, which do not depend on the particular stable magnetic configuration, but again are controlled by the external superconducting phase. The oscillations of individual magnets in these eigenmodes are strongly coupled. The modes can be interpreted as non-degenerated acoustic mode and $N-1$ degenerate optic modes. 

\begin{acknowledgments}
The work has been supported by RSF project No. 22-42-04408. 
\end{acknowledgments}

\bibliography{many_magnets}

\end{document}